\documentclass[superscriptaddress,aps,twocolumn,showpacs]{revtex4-1}
\usepackage{graphicx}
\usepackage{color}
\usepackage{latexsym}

\begin{document}

This manuscript has been authored by UT-Battelle, LLC under Contract No. DE-AC05-00OR22725 with the U.S. Department of Energy. The United States Government retains and the publisher, by accepting the article for publication, acknowledges that the United States Government retains a non-exclusive, paid-up, irrevocable, world-wide license to publish or reproduce the published form of this manuscript, or allow others to do so, for United States Government purposes. The Department of Energy will provide public access to these results of federally sponsored research in accordance with the DOE Public Access Plan (http://energy.gov/downloads/doe-public-access-plan). \clearpage

\title{Antisite pairs suppress the thermal conductivity of BAs}

\author{Qiang Zheng}
\affiliation{Materials Science and Technology Division, Oak Ridge National Laboratory,
Oak Ridge, TN 37831, USA}
\author{Carlos A. Polanco}
\affiliation{Materials Science and Technology Division, Oak Ridge National Laboratory,
Oak Ridge, TN 37831, USA}
\author{Mao-Hua Du}
\affiliation{Materials Science and Technology Division, Oak Ridge National Laboratory,
Oak Ridge, TN 37831, USA}
\author{Lucas R. Lindsay}
\affiliation{Materials Science and Technology Division, Oak Ridge National Laboratory,
Oak Ridge, TN 37831, USA}
\author{Miaofang Chi}
\affiliation{Center for Nanophase Materials Sciences, Oak Ridge National Laboratory,
Oak Ridge, TN 37831, USA}
\author{Jiaqiang Yan}
\affiliation{Materials Science and Technology Division, Oak Ridge National Laboratory,
Oak Ridge, TN 37831, USA}
\affiliation{Department of Materials Science and Engineering, University of Tennessee, Knoxville, TN 37996, USA}
\author{Brian C. Sales}
\affiliation{Materials Science and Technology Division, Oak Ridge National Laboratory,
Oak Ridge, TN 37831, USA}

\date{\today}

\begin{abstract}

BAs was predicted to have an unusually high thermal conductivity at room temperature of 2000\,Wm$^{-1}$\,K$^{-1}$, comparable to that of diamond. However, the experimentally measured thermal conductivity of BAs single crystals is an order of magnitude lower. To identify the origin of this large inconsistency, we investigated the lattice structure and potential defects in BAs single crystals  at atomic scale using aberration-corrected scanning transmission electron microscopy (STEM). Rather than finding a large concentration As vacancies ($V_\mathrm{As}$), as widely thought to dominate the thermal resistance in BAs crystals, our STEM results showed enhanced intensity of some B columns and reduced intensity of some As columns, suggesting the presence of antisite defects with As$_\mathrm{B}$ (As-atom on B site) and B$_\mathrm{As}$ (B-atom on As site) with significant concentrations. Further calculations show that the antisite pair with  As$_\mathrm{B}$ next to B$_\mathrm{As}$ is preferred energetically  among the different types of point defects investigated,
and confirm that such defects lower the thermal conductivity for BAs. Using a concentration of 6.6$\pm$3$\times$10$^{20}$\,cm$^{-3}$ for the antisite pairs estimated from STEM images, thermal conductivity is estimated to be 65--100\,Wm$^{-1}$\,K$^{-1}$, in reasonable agreement with our measured value. Our study suggests that  As$_\mathrm{B}$-B$_\mathrm{As}$ antisite pairs are the primary lattice defects suppressing thermal conductivity of BAs. Possible approaches are proposed for growth of high quality crystals or films with high thermal conductivity.

\end{abstract}


\maketitle

As microelectronic devices develop towards miniaturization and faster processing,
thermal management plays a crucial role in the design of electronics packaging.
Therefore, materials with a high thermal conductivity ($\kappa$) are becoming increasingly essential for new generation electronic devices \cite{MOORE2014}. Recently, cubic boron arsenide (BAs) was predicted to possess an exceptionally high $\kappa$
over 2000\,Wm$^{-1}$\,K$^{-1}$ at room temperature based on first-principles calculations \cite{Lindsay2013PRL,Broido2013PRB}, comparable to that of diamond. The high $\kappa$ in BAs is attributed to the combination of a large acoustic-optic frequency gap and a bunching of the acoustic phonon dispersions, which significantly reduce phonon-phonon scattering \cite{Lindsay2013PRL,Broido2013PRB,ma2016boron}.
This remarkably high $\kappa$ attracted intense attention, however, experimental studies of BAs single crystals found the room temperature $\kappa$ was only $\sim$ 200\,Wm$^{-1}$\,K$^{-1}$ \cite{Kim2016APL,Lv2016APL} and recently improved to be $\sim$ 350\,Wm$^{-1}$\,K$^{-1}$ \cite{Tian2018APL}, an
order of magnitude lower than the predicted value.
Further $ab$ initio calculations suggested that As vacancies, even with a concentration as low as 0.004\,\%, could effectively suppress $\kappa$ \cite{Protik2016PRB}. This seemed to be supported by X-ray photoelectron spectroscopy (XPS) studies which suggested 0.4\,\% \cite{Kim2016APL} or 2.8\,\% \cite{Lv2016APL} As-deficiency in BAs single crystals. Nevertheless, there has been no direct observation of As-vacancies in BAs crystals or films. Identifying the defects that suppress $\kappa$, could provide effective guidance to the growth of defect-free BAs crystals or films with greatly
improved $\kappa$. Since lattice defects are in general quite local, atomic scale scanning transmission electron microscopy (STEM) is a powerful tool for such investigations.

In this Letter, we investigate possible defects in BAs single crystals at atomic scale utilizing aberration-corrected
STEM combined with DFT calculations. As$_\mathrm{B}$-B$_\mathrm{As}$ antisite pairs are identified as the primary lattice defects suppressing the thermal conductivity of BAs. Using a concentration of 6.6$\pm$3$\times$10$^{20}$\,cm$^{-3}$ for the antisite pairs estimated from STEM images,  thermal conductivity is estimated to be 65--100\,Wm$^{-1}$\,K$^{-1}$, comparable to our measured value.

\begin{figure}
\includegraphics[width=0.48\textwidth]{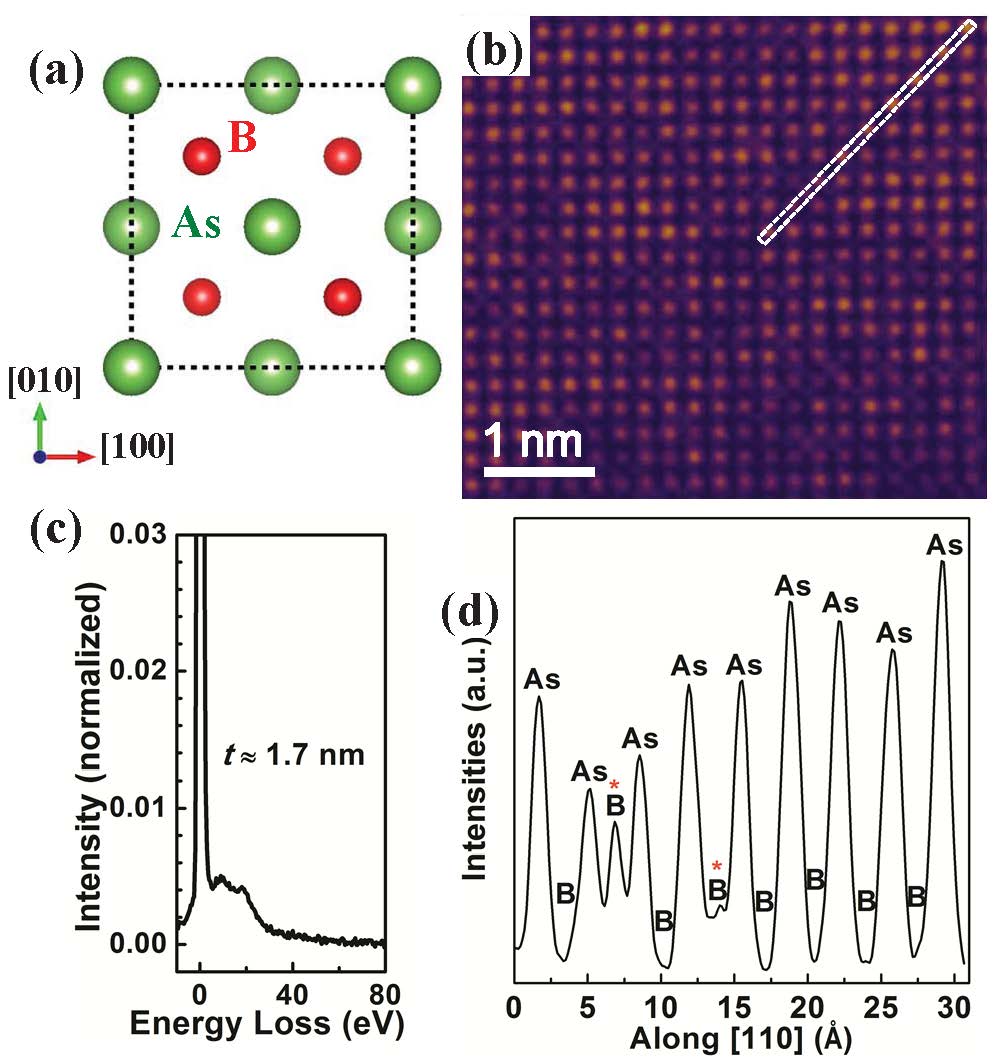}
\caption{(Color online) (a) The crystal structure of BAs in the projection of [001].
Note that each atomic column along this direction  is constructed of atoms of a single type.
(b) A HAADF image along [001] for a region with thickness of 1.7\,nm ($\sim$3.6 unit cells).
The thickness for this region was calculated from its EEL spectrum in (c).
The intensity profile for the dashed rectangular region in (b)  is displayed in (d), revealing As$_\mathrm{B}$ antisite defects and intensity weakening for their neighboring As columns.
The intensities of the two B columns marked by red asterisks in (d) are 0.38 and 0.19, respectively, revealing 1As$_\mathrm{B}$ in each of them, and the intensity difference between them is due to the probe channeling
(see details in Fig.\ S3 in Supplementary Materials \cite{NoteSM}).}
\label{Fig1}
\end{figure}

BAs single crystals were grown by the vapor transport method using iodine
as the transport agent \cite{Chu1972}.
The room temperature $\kappa$ of as-grown single crystals was measured to be $\sim$ 140\,Wm$^{-1}$\,K$^{-1}$ \cite{unpublished},
comparable to the values reported by other groups \cite{Kim2016APL,Lv2016APL}. STEM specimens were prepared by crushing BAs crystals.
The STEM experiments were performed in an aberration-corrected Nion UltraSTEM 100$^\mathrm{TM}$,
operating at 100 kV accelerating voltage \cite{Krivanek2008}. High-angle annular dark-field (HAADF) images were
collected with a probe convergence angle of 30 mrad and an inner collection angle of 86 mrad.
Thickness for each imaging region was measured from the corresponding electron energy loss (EEL) spectrum
using the log-ratio method with an inelastic mean free path ($\lambda$) calculation as described in Refs.\ \cite{Malis1988,egerton2011}.

BAs crystallizes in a zinc blende cubic structure with space group $F\overline{4}$3$m$ and lattice parameter $a$ = 4.7776\,\AA \cite{Perri1958}. As shown in Fig.\ \ref{Fig1}(a), its perfect structure in the projection of [001] reveals
each atomic column involving single type of atoms, either B or As. Fig.\ \ref{Fig1}(b) displays a typical HAADF image along [001] for
a region with thickness of $\sim$1.7\,nm ($\sim$3.6$a$), as measured from the corresponding EEL spectrum (Fig.\ \ref{Fig1}(c)).
In most areas of Fig.\ \ref{Fig1}(b), only As columns show visible intensities, which are
as expected since the HAADF image intensity is roughly proportional to $Z^2$ ($Z$ is atomic number)
and thus in the perfect structure the intensities of B columns should be negligible compared to those of As columns.
However, obvious intensities for some B columns and intensity weakening for some As columns are observed as highlighted by the intensity profile in Fig.\ \ref{Fig1}(d), indicating the appearance of As-atoms on B sites (As$_\mathrm{B}$ antisites) and B-atoms on As sites (B$_\mathrm{As}$ antisites) or As vacancies (V$_\mathrm{As}$). Note that crystals were grown starting with high-purity As and B materials and
no foreign atoms were observed by elemental analysis
and EEL spectroscopy measurements for the defected areas (see Fig.\ S1  \cite{NoteSM}).
To assess the possible effects of the crystal edge, a much thicker region (with thickness of $\sim$4.7\,nm) far away from the edge was chosen for HAADF imaging (see Fig.\ S2 \cite{NoteSM}) and similar features were also observed.

Image simulations were then performed to better visualize the intensity variations caused by possible
As$_\mathrm{B}$ in B columns and B$_\mathrm{As}$ or V$_\mathrm{As}$ in As columns (See Supplementary materials for more details \cite {NoteSM}).
Simulations of the intensity profile with the intensity of a pristine As column normalized to 1,
suggest that the intensity of a B column with 1As$_\mathrm{B}$ could vary between 0.18 and 0.4,
while that of an As column with 1B$_\mathrm{As}$ or 1V$_\mathrm{As}$ could be in the range of  0.55 and 0.94, due to probe channeling \cite{esser2016}.
We then measured intensities for the visible B columns in Fig.\ \ref{Fig1}(b) and other images with same thicknesses.
As exemplified by two B columns containing As$_\mathrm{B}$ antisite defects marked by red asterisks in Fig.\ \ref{Fig1}(d),
intensities of almost all visible B columns are between 0.18 and 0.4, much weaker than the simulated intensities for a B column with 2As$_\mathrm{B}$ (0.69--0.84) or 3As$_\mathrm{B}$ (0.87--0.93) (see Fig.\ S4 \cite{NoteSM}),
suggesting only one As$_\mathrm{B}$ antisite defect in each of them.
By counting the As$_\mathrm{B}$ defects in over ten HAADF images of different regions with comparable thicknesses as in Fig.\ \ref{Fig1}(b),
the concentration was estimated to be 1.8(8)\,\% (6.6$\pm$3$\times$10$^{20}$\,cm$^{-3}$).
The large error bar comes from the fact that the concentration estimation at very thin regions is affected by various facts such as the type of atoms on the top and bottom surfaces.

Although the HAADF imaging and simulations strongly suggest the presence of As$_\mathrm{B}$ antisite defects in BAs, we cannot distinguish the origin of the reduced intensity on the As columns directly,
whether the intensity drop is due to B$_\mathrm{As}$ or V$_\mathrm{As}$.
We thus carefully analyzed the local area variation in HAADF images. Our analysis indicates no obvious lattice expansion or contraction,
suggesting B-atoms on As sites rather than the vacancies produce the intensity features. Meanwhile, as shown in Fig.\ \ref{Fig1}(b) and (d),
B columns with enhanced intensities are always found neighboring to As columns with reduced intensities, indicating pairing of B$_\mathrm{As}$ and As$_\mathrm{B}$ antisite defects in the structure.
We also performed a careful search of other types of defects, with special attention paid to As vacancies, which have been widely believed to suppress $\kappa$ of BAs. However, we did not find any trace of other types of defects with comparable concentration to that of the B$_\mathrm{As}$-As$_\mathrm{B}$ pairs.
Defect formation energy calculations discussed below show B$_\mathrm{As}$-As$_\mathrm{B}$ pairs are the most energetically preferred.

\begin{figure}[!t]
\includegraphics[width=0.45\textwidth]{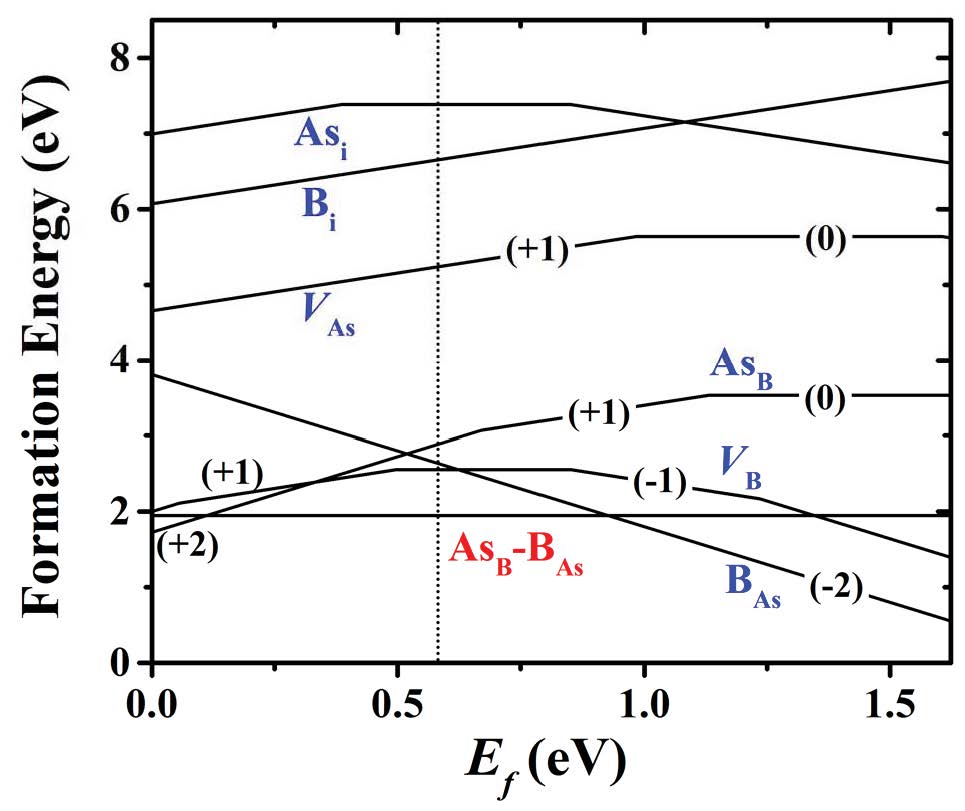}
\caption{(Color online) Formation energies of native point defects (including vacancies, interstitials, and antisites; the red label is for the B$_\mathrm{As}$-As$_\mathrm{B}$ pair) in BAs.
The slope of the formation energy line indicates the charge state of the defect.
The Fermi level is pinned approximately at the crossing point between the formation energy lines of the lowest-energy donor ($V_\mathrm{B}^{+}$) and acceptor (B$_\mathrm{As}^{2-}$) defects (indicated by the dotted line).}
\label{Formation_energies}
\end{figure}

To further understand the origin of this particular defect type in BAs, the formation energies of different types of defects
are calculated (See Supplementary Materials for detailed computational methods \cite{NoteSM}).
Fig.\ \ref{Formation_energies} shows the calculated formation energies of native point defects (vacancies, interstitials, and antisites) in BAs.
The Fermi level is pinned approximately at the crossing point between the formation energy lines of the lowest-energy donor ($V_\mathrm{B}^{+}$) and acceptor (B$_\mathrm{As}^{2-}$) defects
(indicated by the vertical dotted line in Fig.\ \ref{Formation_energies}). At this Fermi level, the antisite pair, As$_\mathrm{B}$-B$_\mathrm{As}$,
has the lowest formation energy (1.95\,eV) among all native point defects, consistent with the STEM result. However, the calculated thermal-equilibrium concentration of As$_\mathrm{B}$-B$_\mathrm{As}$ (on the order of 10$^{13}$ cm$^{-3}$)
at the growth temperature of 850\,$^\mathrm{o}$C (using Eq. (5) in Supplementary materials \cite {NoteSM}) is significantly lower than that observed in HAADF images. The high defect formation energies indicate strong covalent bonding in BAs, which is consistent with the predicted high thermal conductivity.
The high concentration of antisite defects as seen in HAADF images is likely because thermal equilibrium is not reached during crystal growth.
The gas-phase species react to form solid-state BAs in the vapor transport synthesis. It is likely that a large number of antisite pairs are trapped in the crystal lattice.
Thermal annealing is supposed to reduce the defect concentration to its thermal-equilibrium value provided that sufficient atomic diffusion can take place.
However, the atomic diffusion in BAs is likely limited especially for As even at the growth temperature of 850\,$^\mathrm{o}$C for the following reasons:
(1) The As interstitial (As$_{i}$) and As vacancy ($V_\mathrm{As}$) defects both have very high formation energies as shown in Fig.\ \ref{Formation_energies}.
Thus, the concentrations of As$_{i}$ and $V_\mathrm{As}$ are likely orders of magnitudes lower than that of As$_\mathrm{B}$-B$_\mathrm{As}$ regardless of whether thermal equilibrium can be reached.
(2) BAs has a small lattice constant but a large size mismatch between B and As. As a result, the diffusion of the large As$_{i}$ interstitial is likely difficult.
The diffusion of $V_\mathrm{As}$ involves creating two $V_\mathrm{As}$ and one As$_{i}$ at the transition state of the diffusion path,
which may lead to a high diffusion barrier because both $V_\mathrm{As}$ and As$_{i}$ have very high formation energies.
Thus, the low concentration and high diffusion barrier of As$_{i}$ and $V_\mathrm{As}$ may severely limit As diffusion in BAs.
This prevents the As$_\mathrm{B}$-B$_\mathrm{As}$ defect from reaching its thermal equilibrium; thereby, trapping a substantial amount of As$_\mathrm{B}$-B$_\mathrm{As}$ defects in BAs as seen in HAADF images.

\begin{figure}[!t]
\includegraphics[width=0.45\textwidth]{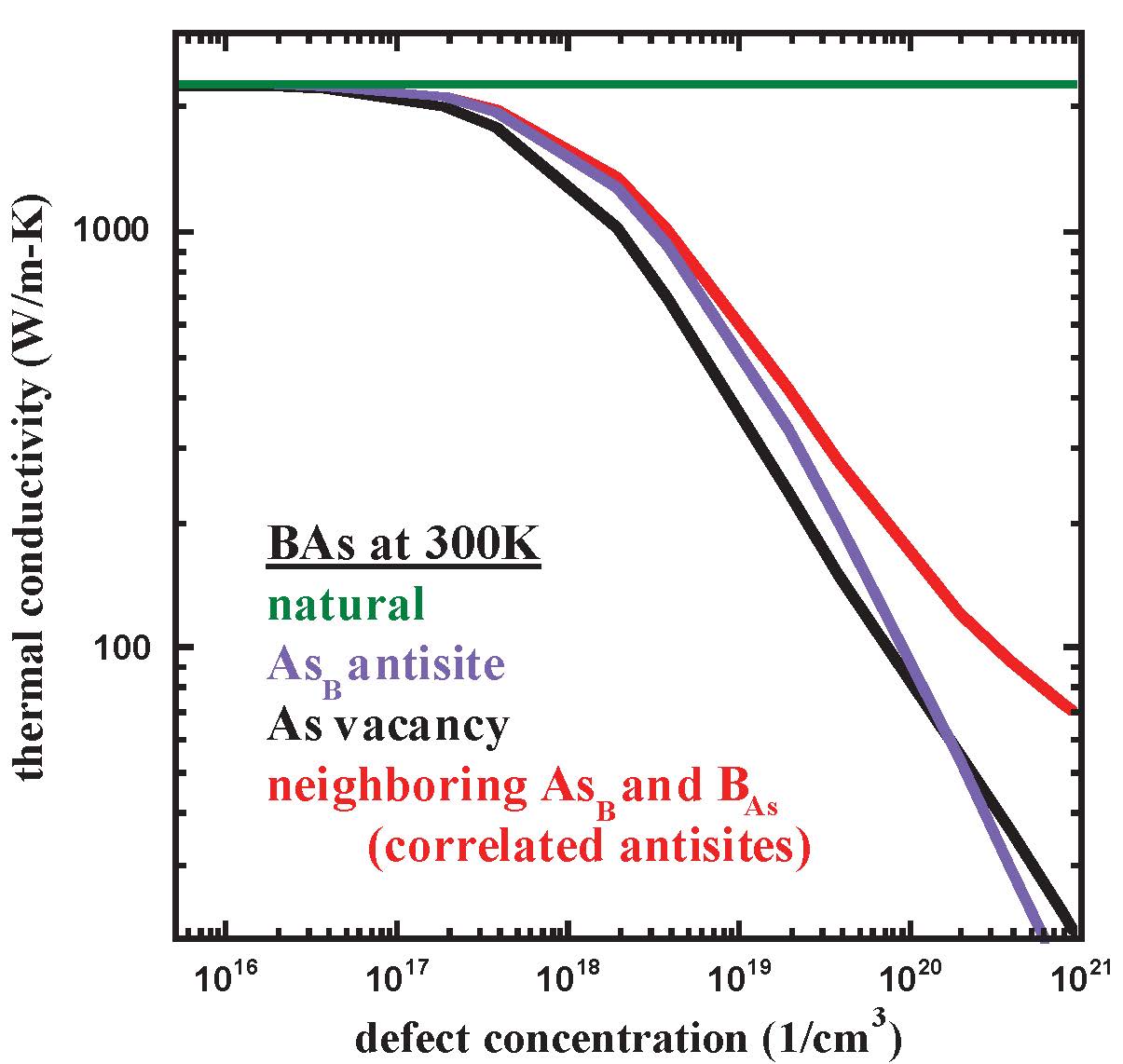}
\caption{(Color online) Thermal conductivity ($\kappa$) of BAs at room temperature for
As$_\mathrm{B}$ antisite defects (purple curve), As vacancies (black curve), and neighboring
As$_\mathrm{B}$ and B$_\mathrm{As}$ antisite defects (red curve) as a function of defect concentration.
The horizontal green line gives the room temperature calculated $\kappa$ for BAs with natural isotope variation.}
\label{Kappa}
\end{figure}

 Extrinsic thermal resistance in a material with extended and point defects becomes significant as intrinsic anharmonic resistance becomes weak,
 for example with decreasing temperature.  For high thermal conductivity ($\kappa$) materials this can be exaggerated,
 as is the case for diamond and graphene where phonon-isotope scattering has been shown to reduce their $\kappa$ by more than 50\,\% even at RT \cite{Anthony1990,Chen2012}.
 In BAs with predicted $\kappa$ \textgreater 2000\,Wm$^{-1}$\,K$^{-1}$ \cite{Lindsay2013PRL} phonon-defect scattering may also be extremely important,
 especially in validating the prediction by experiment. Previous theoretical work demonstrated that 0.004\,\% As vacancies ($\sim$1.5$\times$10$^{18}$ cm$^{-3}$) reduces the predicted $\kappa$ by half \cite{Protik2016PRB}.
 Thus, the large concentration of antisite defects observed here are likely a leading factor in the much reduced thermal conductivity observed experimentally, $\kappa$  $\sim$140\,Wm$^{-1}$\,K$^{-1}$ \cite{unpublished}.
 Fig.\ \ref{Kappa}  shows $\kappa$ of BAs calculated using the full solution of the Peierls-Boltzmann transport equation with first-principles interatomic force constants \cite{Ziman2001,Srivastava1990,Lindsay2013PRB}. A parameter-free $ab$ initio Green's function methodology \cite{Mingo2010,Protik2016PRB,Polanco2018}, which has demonstrated good agreement with measured $\kappa$ data \cite{Katcho2014,Katre2017}, was used to include phonon-defect scattering from different defect types with varying concentration (See Supplementary materials \cite {NoteSM}).
 Using the estimated concentration of As$_\mathrm{B}$-B$_\mathrm{As}$ pairs from the STEM measurements here (6.6$\pm$3$\times$10$^{20}$\,cm$^{-3}$), calculations give $\kappa$ 65--100\,Wm$^{-1}$\,K$^{-1}$, comparable to  the measured value.

 The identification of As$_\mathrm{B}$-B$_\mathrm{As}$ pairs as the primary defects suppressing  $\kappa$ of BAs and the high formation energy from DFT calculations provide important information and highlight the importance of kinetic factors during synthesizing high quality BAs materials with predicted high $\kappa$.
 Tuning the pressure and/or temperature might change both the chemical potential of vapor species inside of the growth ampoule and the growth kinetics in vapor transport synthesis. Growth of BAs crystals out of flux might be a more promising approach, though challenging due to the limited solubility of B in most low melting fluxes. A thorough investigation of phase diagrams suggests Ni- or alkali metals-based fluxes are promising with reasonable solubility of B \cite{Portnoi1967,naslain1970,Borgstedt2003,Kang2017}. Considering the growth of B$_{12}$As$_2$ out of NiB melt \cite{Whiteley2011}, the B content in the Ni-based flux should be carefully controlled and an As-rich Ni-based flux is recommended to avoid the precipitation of B$_{12}$As$_2$. For the growth of high quality BAs films, molecular beam epitaxy (MBE) might be a good option.

In summary, with a combined effort of STEM imaging and DFT calculations, we identify As$_\mathrm{B}$-B$_\mathrm{As}$ antisite pairs are the primary lattice defects rather than As vacancies suppressing thermal conductivity of BAs single crystals. Further studies are needed to understand the kinetic factors leading to the formation of these lattice defects during vapor transport growth. Flux growth out of alkali metals-based or Ni-based melts might be a good option for high quality crystals. Considering the sensitivity of thermal conductivity to lattice defects, MBE is suggested for the growth of BAs films.

This work was supported by the U.S. Department of Energy (DOE), Office of Science, Basic Energy Sciences (BES), Materials Sciences and Engineering Division.
The electron microscopy in this work was conducted at the ORNL's Center for Nanophase Materials Sciences (CNMS), which is a DOE Office of Science User Facility.
C.A.P. and L.L. acknowledge computational resources from the National Energy Research Scientific Computing Center (NERSC),
a DOE Office of Science User Facility supported by the Office of Science of the US Department of Energy under Contract No. DE-AC02-05CH11231.


\begin{thebibliography}{30}%
\makeatletter
\providecommand \@ifxundefined [1]{%
 \@ifx{#1\undefined}
}%
\providecommand \@ifnum [1]{%
 \ifnum #1\expandafter \@firstoftwo
 \else \expandafter \@secondoftwo
 \fi
}%
\providecommand \@ifx [1]{%
 \ifx #1\expandafter \@firstoftwo
 \else \expandafter \@secondoftwo
 \fi
}%
\providecommand \natexlab [1]{#1}%
\providecommand \enquote  [1]{``#1''}%
\providecommand \bibnamefont  [1]{#1}%
\providecommand \bibfnamefont [1]{#1}%
\providecommand \citenamefont [1]{#1}%
\providecommand \href@noop [0]{\@secondoftwo}%
\providecommand \href [0]{\begingroup \@sanitize@url \@href}%
\providecommand \@href[1]{\@@startlink{#1}\@@href}%
\providecommand \@@href[1]{\endgroup#1\@@endlink}%
\providecommand \@sanitize@url [0]{\catcode `\\12\catcode `\$12\catcode
  `\&12\catcode `\#12\catcode `\^12\catcode `\_12\catcode `\%12\relax}%
\providecommand \@@startlink[1]{}%
\providecommand \@@endlink[0]{}%
\providecommand \url  [0]{\begingroup\@sanitize@url \@url }%
\providecommand \@url [1]{\endgroup\@href {#1}{\urlprefix }}%
\providecommand \urlprefix  [0]{URL }%
\providecommand \Eprint [0]{\href }%
\providecommand \doibase [0]{http://dx.doi.org/}%
\providecommand \selectlanguage [0]{\@gobble}%
\providecommand \bibinfo  [0]{\@secondoftwo}%
\providecommand \bibfield  [0]{\@secondoftwo}%
\providecommand \translation [1]{[#1]}%
\providecommand \BibitemOpen [0]{}%
\providecommand \bibitemStop [0]{}%
\providecommand \bibitemNoStop [0]{.\EOS\space}%
\providecommand \EOS [0]{\spacefactor3000\relax}%
\providecommand \BibitemShut  [1]{\csname bibitem#1\endcsname}%
\let\auto@bib@innerbib\@empty
\bibitem [{\citenamefont {Moore}\ and\ \citenamefont {Shi}(2014)}]{MOORE2014}%
  \BibitemOpen
  \bibfield  {author} {\bibinfo {author} {\bibfnamefont {A.~L.}\ \bibnamefont
  {Moore}}\ and\ \bibinfo {author} {\bibfnamefont {L.}~\bibnamefont {Shi}},\
  }\href {\doibase 10.1016/j.mattod.2014.04.003} {\bibfield  {journal}
  {\bibinfo  {journal} {Materials Today}\ }\textbf {\bibinfo {volume} {17}},\
  \bibinfo {pages} {163} (\bibinfo {year} {2014})}\BibitemShut {NoStop}%
\bibitem [{\citenamefont {Lindsay}\ \emph
  {et~al.}(2013{\natexlab{a}})\citenamefont {Lindsay}, \citenamefont {Broido},\
  and\ \citenamefont {Reinecke}}]{Lindsay2013PRL}%
  \BibitemOpen
  \bibfield  {author} {\bibinfo {author} {\bibfnamefont {L.}~\bibnamefont
  {Lindsay}}, \bibinfo {author} {\bibfnamefont {D.~A.}\ \bibnamefont {Broido}},
  \ and\ \bibinfo {author} {\bibfnamefont {T.~L.}\ \bibnamefont {Reinecke}},\
  }\href {\doibase 10.1103/PhysRevLett.111.025901} {\bibfield  {journal}
  {\bibinfo  {journal} {Phys. Rev. Lett.}\ }\textbf {\bibinfo {volume} {111}},\
  \bibinfo {pages} {025901} (\bibinfo {year} {2013}{\natexlab{a}})}\BibitemShut
  {NoStop}%
\bibitem [{\citenamefont {Broido}\ \emph {et~al.}(2013)\citenamefont {Broido},
  \citenamefont {Lindsay},\ and\ \citenamefont {Reinecke}}]{Broido2013PRB}%
  \BibitemOpen
  \bibfield  {author} {\bibinfo {author} {\bibfnamefont {D.~A.}\ \bibnamefont
  {Broido}}, \bibinfo {author} {\bibfnamefont {L.}~\bibnamefont {Lindsay}}, \
  and\ \bibinfo {author} {\bibfnamefont {T.~L.}\ \bibnamefont {Reinecke}},\
  }\href {\doibase 10.1103/PhysRevB.88.214303} {\bibfield  {journal} {\bibinfo
  {journal} {Phys. Rev. B}\ }\textbf {\bibinfo {volume} {88}},\ \bibinfo
  {pages} {214303} (\bibinfo {year} {2013})}\BibitemShut {NoStop}%
\bibitem [{\citenamefont {Ma}\ \emph {et~al.}(2016)\citenamefont {Ma},
  \citenamefont {Li}, \citenamefont {Tang}, \citenamefont {Yan}, \citenamefont
  {Alatas}, \citenamefont {Lindsay}, \citenamefont {Sales},\ and\ \citenamefont
  {Tian}}]{ma2016boron}%
  \BibitemOpen
  \bibfield  {author} {\bibinfo {author} {\bibfnamefont {H.}~\bibnamefont
  {Ma}}, \bibinfo {author} {\bibfnamefont {C.}~\bibnamefont {Li}}, \bibinfo
  {author} {\bibfnamefont {S.}~\bibnamefont {Tang}}, \bibinfo {author}
  {\bibfnamefont {J.}~\bibnamefont {Yan}}, \bibinfo {author} {\bibfnamefont
  {A.}~\bibnamefont {Alatas}}, \bibinfo {author} {\bibfnamefont
  {L.}~\bibnamefont {Lindsay}}, \bibinfo {author} {\bibfnamefont {B.~C.}\
  \bibnamefont {Sales}}, \ and\ \bibinfo {author} {\bibfnamefont
  {Z.}~\bibnamefont {Tian}},\ }\href@noop {} {\bibfield  {journal} {\bibinfo
  {journal} {Phys.\ Rev.\ B}\ }\textbf {\bibinfo {volume} {94}},\ \bibinfo
  {pages} {220303} (\bibinfo {year} {2016})}\BibitemShut {NoStop}%
\bibitem [{\citenamefont {Kim}\ \emph {et~al.}(2016)\citenamefont {Kim},
  \citenamefont {Evans}, \citenamefont {Sellan}, \citenamefont {Williams},
  \citenamefont {Ou}, \citenamefont {Cowley},\ and\ \citenamefont
  {Shi}}]{Kim2016APL}%
  \BibitemOpen
  \bibfield  {author} {\bibinfo {author} {\bibfnamefont {J.}~\bibnamefont
  {Kim}}, \bibinfo {author} {\bibfnamefont {D.~A.}\ \bibnamefont {Evans}},
  \bibinfo {author} {\bibfnamefont {D.~P.}\ \bibnamefont {Sellan}}, \bibinfo
  {author} {\bibfnamefont {O.~M.}\ \bibnamefont {Williams}}, \bibinfo {author}
  {\bibfnamefont {E.}~\bibnamefont {Ou}}, \bibinfo {author} {\bibfnamefont
  {A.~H.}\ \bibnamefont {Cowley}}, \ and\ \bibinfo {author} {\bibfnamefont
  {L.}~\bibnamefont {Shi}},\ }\href {\doibase 10.1063/1.4950970} {\bibfield
  {journal} {\bibinfo  {journal} {Applied Physics Letters}\ }\textbf {\bibinfo
  {volume} {108}},\ \bibinfo {pages} {201905} (\bibinfo {year} {2016})},\
  \Eprint {http://arxiv.org/abs/https://doi.org/10.1063/1.4950970}
  {https://doi.org/10.1063/1.4950970} \BibitemShut {NoStop}%
\bibitem [{\citenamefont {Lv}\ \emph {et~al.}(2015)\citenamefont {Lv},
  \citenamefont {Lan}, \citenamefont {Wang}, \citenamefont {Zhang},
  \citenamefont {Hu}, \citenamefont {Jacobson}, \citenamefont {Broido},
  \citenamefont {Chen}, \citenamefont {Ren},\ and\ \citenamefont
  {Chu}}]{Lv2016APL}%
  \BibitemOpen
  \bibfield  {author} {\bibinfo {author} {\bibfnamefont {B.}~\bibnamefont
  {Lv}}, \bibinfo {author} {\bibfnamefont {Y.}~\bibnamefont {Lan}}, \bibinfo
  {author} {\bibfnamefont {X.}~\bibnamefont {Wang}}, \bibinfo {author}
  {\bibfnamefont {Q.}~\bibnamefont {Zhang}}, \bibinfo {author} {\bibfnamefont
  {Y.}~\bibnamefont {Hu}}, \bibinfo {author} {\bibfnamefont {A.~J.}\
  \bibnamefont {Jacobson}}, \bibinfo {author} {\bibfnamefont {D.}~\bibnamefont
  {Broido}}, \bibinfo {author} {\bibfnamefont {G.}~\bibnamefont {Chen}},
  \bibinfo {author} {\bibfnamefont {Z.}~\bibnamefont {Ren}}, \ and\ \bibinfo
  {author} {\bibfnamefont {C.-W.}\ \bibnamefont {Chu}},\ }\href {\doibase
  10.1063/1.4913441} {\bibfield  {journal} {\bibinfo  {journal} {Applied
  Physics Letters}\ }\textbf {\bibinfo {volume} {106}},\ \bibinfo {pages}
  {074105} (\bibinfo {year} {2015})},\ \Eprint
  {http://arxiv.org/abs/https://doi.org/10.1063/1.4913441}
  {https://doi.org/10.1063/1.4913441} \BibitemShut {NoStop}%
\bibitem [{\citenamefont {Tian}\ \emph {et~al.}(2018)\citenamefont {Tian},
  \citenamefont {Song}, \citenamefont {Lv}, \citenamefont {Sun}, \citenamefont
  {Huyan}, \citenamefont {Wu}, \citenamefont {Mao}, \citenamefont {Ni},
  \citenamefont {Ding}, \citenamefont {Huberman}, \citenamefont {Liu},
  \citenamefont {Chen}, \citenamefont {Chen}, \citenamefont {Chu},\ and\
  \citenamefont {Ren}}]{Tian2018APL}%
  \BibitemOpen
  \bibfield  {author} {\bibinfo {author} {\bibfnamefont {F.}~\bibnamefont
  {Tian}}, \bibinfo {author} {\bibfnamefont {B.}~\bibnamefont {Song}}, \bibinfo
  {author} {\bibfnamefont {B.}~\bibnamefont {Lv}}, \bibinfo {author}
  {\bibfnamefont {J.}~\bibnamefont {Sun}}, \bibinfo {author} {\bibfnamefont
  {S.}~\bibnamefont {Huyan}}, \bibinfo {author} {\bibfnamefont
  {Q.}~\bibnamefont {Wu}}, \bibinfo {author} {\bibfnamefont {J.}~\bibnamefont
  {Mao}}, \bibinfo {author} {\bibfnamefont {Y.}~\bibnamefont {Ni}}, \bibinfo
  {author} {\bibfnamefont {Z.}~\bibnamefont {Ding}}, \bibinfo {author}
  {\bibfnamefont {S.}~\bibnamefont {Huberman}}, \bibinfo {author}
  {\bibfnamefont {T.-H.}\ \bibnamefont {Liu}}, \bibinfo {author} {\bibfnamefont
  {G.}~\bibnamefont {Chen}}, \bibinfo {author} {\bibfnamefont {S.}~\bibnamefont
  {Chen}}, \bibinfo {author} {\bibfnamefont {C.-W.}\ \bibnamefont {Chu}}, \
  and\ \bibinfo {author} {\bibfnamefont {Z.}~\bibnamefont {Ren}},\ }\href
  {\doibase 10.1063/1.5004200} {\bibfield  {journal} {\bibinfo  {journal}
  {Applied Physics Letters}\ }\textbf {\bibinfo {volume} {112}},\ \bibinfo
  {pages} {031903} (\bibinfo {year} {2018})}\BibitemShut {NoStop}%
\bibitem [{\citenamefont {Protik}\ \emph {et~al.}(2016)\citenamefont {Protik},
  \citenamefont {Carrete}, \citenamefont {Katcho}, \citenamefont {Mingo},\ and\
  \citenamefont {Broido}}]{Protik2016PRB}%
  \BibitemOpen
  \bibfield  {author} {\bibinfo {author} {\bibfnamefont {N.~H.}\ \bibnamefont
  {Protik}}, \bibinfo {author} {\bibfnamefont {J.}~\bibnamefont {Carrete}},
  \bibinfo {author} {\bibfnamefont {N.~A.}\ \bibnamefont {Katcho}}, \bibinfo
  {author} {\bibfnamefont {N.}~\bibnamefont {Mingo}}, \ and\ \bibinfo {author}
  {\bibfnamefont {D.}~\bibnamefont {Broido}},\ }\href {\doibase
  10.1103/PhysRevB.94.045207} {\bibfield  {journal} {\bibinfo  {journal} {Phys.
  Rev. B}\ }\textbf {\bibinfo {volume} {94}},\ \bibinfo {pages} {045207}
  (\bibinfo {year} {2016})}\BibitemShut {NoStop}%
\bibitem [{Not()}]{NoteSM}%
  \BibitemOpen
  \href@noop {} {}\bibinfo {note} {See Supplemental Material at [URL will be
  inserted by publisher]}\BibitemShut {NoStop}%
\bibitem [{\citenamefont {Chu}\ and\ \citenamefont {Hyslop}(1972)}]{Chu1972}%
  \BibitemOpen
  \bibfield  {author} {\bibinfo {author} {\bibfnamefont {T.~L.}\ \bibnamefont
  {Chu}}\ and\ \bibinfo {author} {\bibfnamefont {A.~E.}\ \bibnamefont
  {Hyslop}},\ }\href {\doibase 10.1063/1.1661106} {\bibfield  {journal}
  {\bibinfo  {journal} {Journal of Applied Physics}\ }\textbf {\bibinfo
  {volume} {43}},\ \bibinfo {pages} {276} (\bibinfo {year} {1972})}\BibitemShut
  {NoStop}%
\bibitem [{\citenamefont {Yan}()}]{unpublished}%
  \BibitemOpen
  \bibfield  {author} {\bibinfo {author} {\bibfnamefont {J.-Q.}\ \bibnamefont
  {Yan}},\ }\href@noop {} {}\bibinfo {note} {Unpublished Data}\BibitemShut
  {NoStop}%
\bibitem [{\citenamefont {Krivanek}\ \emph {et~al.}(2008)\citenamefont
  {Krivanek}, \citenamefont {Corbin}, \citenamefont {Dellby}, \citenamefont
  {Elston}, \citenamefont {Keyse}, \citenamefont {Murfitt}, \citenamefont
  {Own}, \citenamefont {Szilagyi},\ and\ \citenamefont
  {Woodruff}}]{Krivanek2008}%
  \BibitemOpen
  \bibfield  {author} {\bibinfo {author} {\bibfnamefont {O.}~\bibnamefont
  {Krivanek}}, \bibinfo {author} {\bibfnamefont {G.}~\bibnamefont {Corbin}},
  \bibinfo {author} {\bibfnamefont {N.}~\bibnamefont {Dellby}}, \bibinfo
  {author} {\bibfnamefont {B.}~\bibnamefont {Elston}}, \bibinfo {author}
  {\bibfnamefont {R.}~\bibnamefont {Keyse}}, \bibinfo {author} {\bibfnamefont
  {M.}~\bibnamefont {Murfitt}}, \bibinfo {author} {\bibfnamefont
  {C.}~\bibnamefont {Own}}, \bibinfo {author} {\bibfnamefont {Z.}~\bibnamefont
  {Szilagyi}}, \ and\ \bibinfo {author} {\bibfnamefont {J.}~\bibnamefont
  {Woodruff}},\ }\href@noop {} {\bibfield  {journal} {\bibinfo  {journal}
  {Ultramicroscopy}\ }\textbf {\bibinfo {volume} {108}},\ \bibinfo {pages}
  {179} (\bibinfo {year} {2008})}\BibitemShut {NoStop}%
\bibitem [{\citenamefont {Malis}\ \emph {et~al.}(1988)\citenamefont {Malis},
  \citenamefont {Cheng},\ and\ \citenamefont {Egerton}}]{Malis1988}%
  \BibitemOpen
  \bibfield  {author} {\bibinfo {author} {\bibfnamefont {T.}~\bibnamefont
  {Malis}}, \bibinfo {author} {\bibfnamefont {S.~C.}\ \bibnamefont {Cheng}}, \
  and\ \bibinfo {author} {\bibfnamefont {R.~F.}\ \bibnamefont {Egerton}},\
  }\href@noop {} {\bibfield  {journal} {\bibinfo  {journal} {Microsc.\ Res.\
  Tech.}\ }\textbf {\bibinfo {volume} {8}},\ \bibinfo {pages} {193} (\bibinfo
  {year} {1988})}\BibitemShut {NoStop}%
\bibitem [{\citenamefont {Egerton}(2011)}]{egerton2011}%
  \BibitemOpen
  \bibfield  {author} {\bibinfo {author} {\bibfnamefont {R.}~\bibnamefont
  {Egerton}},\ }\href@noop {} {\emph {\bibinfo {title} {Electron energy-loss
  spectroscopy in the electron microscope}}}\ (\bibinfo  {publisher} {Springer
  Science \& Business Media},\ \bibinfo {year} {2011})\BibitemShut {NoStop}%
\bibitem [{\citenamefont {Perri}\ \emph {et~al.}(1958)\citenamefont {Perri},
  \citenamefont {La~Placa},\ and\ \citenamefont {Post}}]{Perri1958}%
  \BibitemOpen
  \bibfield  {author} {\bibinfo {author} {\bibfnamefont {J.~A.}\ \bibnamefont
  {Perri}}, \bibinfo {author} {\bibfnamefont {S.}~\bibnamefont {La~Placa}}, \
  and\ \bibinfo {author} {\bibfnamefont {B.}~\bibnamefont {Post}},\ }\href
  {\doibase 10.1107/S0365110X58000827} {\bibfield  {journal} {\bibinfo
  {journal} {Acta Crystallographica}\ }\textbf {\bibinfo {volume} {4}},\
  \bibinfo {pages} {310} (\bibinfo {year} {1958})}\BibitemShut {NoStop}%
\bibitem [{\citenamefont {Esser}\ \emph {et~al.}(2016)\citenamefont {Esser},
  \citenamefont {Hauser}, \citenamefont {Williams}, \citenamefont {Allen},
  \citenamefont {Woodward}, \citenamefont {Yang},\ and\ \citenamefont
  {McComb}}]{esser2016}%
  \BibitemOpen
  \bibfield  {author} {\bibinfo {author} {\bibfnamefont {B.~D.}\ \bibnamefont
  {Esser}}, \bibinfo {author} {\bibfnamefont {A.~J.}\ \bibnamefont {Hauser}},
  \bibinfo {author} {\bibfnamefont {R.~E.~A.}\ \bibnamefont {Williams}},
  \bibinfo {author} {\bibfnamefont {L.~J.}\ \bibnamefont {Allen}}, \bibinfo
  {author} {\bibfnamefont {P.~M.}\ \bibnamefont {Woodward}}, \bibinfo {author}
  {\bibfnamefont {F.~Y.}\ \bibnamefont {Yang}}, \ and\ \bibinfo {author}
  {\bibfnamefont {D.~W.}\ \bibnamefont {McComb}},\ }\href@noop {} {\bibfield
  {journal} {\bibinfo  {journal} {Phys.\ Rev.\ Lett.}\ }\textbf {\bibinfo
  {volume} {117}},\ \bibinfo {pages} {176101} (\bibinfo {year}
  {2016})}\BibitemShut {NoStop}%
\bibitem [{\citenamefont {Anthony}\ \emph {et~al.}(1990)\citenamefont
  {Anthony}, \citenamefont {Banholzer}, \citenamefont {Fleischer},
  \citenamefont {Wei}, \citenamefont {Kuo}, \citenamefont {Thomas},\ and\
  \citenamefont {Pryor}}]{Anthony1990}%
  \BibitemOpen
  \bibfield  {author} {\bibinfo {author} {\bibfnamefont {T.~R.}\ \bibnamefont
  {Anthony}}, \bibinfo {author} {\bibfnamefont {W.~F.}\ \bibnamefont
  {Banholzer}}, \bibinfo {author} {\bibfnamefont {J.~F.}\ \bibnamefont
  {Fleischer}}, \bibinfo {author} {\bibfnamefont {L.}~\bibnamefont {Wei}},
  \bibinfo {author} {\bibfnamefont {P.~K.}\ \bibnamefont {Kuo}}, \bibinfo
  {author} {\bibfnamefont {R.~L.}\ \bibnamefont {Thomas}}, \ and\ \bibinfo
  {author} {\bibfnamefont {R.~W.}\ \bibnamefont {Pryor}},\ }\href {\doibase
  10.1103/PhysRevB.42.1104} {\bibfield  {journal} {\bibinfo  {journal} {Phys.
  Rev. B}\ }\textbf {\bibinfo {volume} {42}},\ \bibinfo {pages} {1104}
  (\bibinfo {year} {1990})}\BibitemShut {NoStop}%
\bibitem [{\citenamefont {Chen}\ \emph {et~al.}(2012)\citenamefont {Chen},
  \citenamefont {Wu}, \citenamefont {Mishra}, \citenamefont {Kang},
  \citenamefont {Zhang}, \citenamefont {Cho}, \citenamefont {Cai},
  \citenamefont {Balandin},\ and\ \citenamefont {Ruoff}}]{Chen2012}%
  \BibitemOpen
  \bibfield  {author} {\bibinfo {author} {\bibfnamefont {S.}~\bibnamefont
  {Chen}}, \bibinfo {author} {\bibfnamefont {Q.}~\bibnamefont {Wu}}, \bibinfo
  {author} {\bibfnamefont {C.}~\bibnamefont {Mishra}}, \bibinfo {author}
  {\bibfnamefont {J.}~\bibnamefont {Kang}}, \bibinfo {author} {\bibfnamefont
  {H.}~\bibnamefont {Zhang}}, \bibinfo {author} {\bibfnamefont
  {K.}~\bibnamefont {Cho}}, \bibinfo {author} {\bibfnamefont {W.}~\bibnamefont
  {Cai}}, \bibinfo {author} {\bibfnamefont {A.~A.}\ \bibnamefont {Balandin}}, \
  and\ \bibinfo {author} {\bibfnamefont {R.~S.}\ \bibnamefont {Ruoff}},\
  }\href@noop {} {\bibfield  {journal} {\bibinfo  {journal} {Nature Materials}\
  }\textbf {\bibinfo {volume} {11}},\ \bibinfo {pages} {203} (\bibinfo {year}
  {2012})}\BibitemShut {NoStop}%
\bibitem [{\citenamefont {Ziman}(2001)}]{Ziman2001}%
  \BibitemOpen
  \bibfield  {author} {\bibinfo {author} {\bibfnamefont {J.~M.}\ \bibnamefont
  {Ziman}},\ }\href@noop {} {\emph {\bibinfo {title} {Electrons and Phonons:
  The Theory of Transport Phenomena in Solids}}}\ (\bibinfo  {publisher}
  {Clarendon Press},\ \bibinfo {address} {Oxford},\ \bibinfo {year}
  {2001})\BibitemShut {NoStop}%
\bibitem [{\citenamefont {Srivastava}(1990)}]{Srivastava1990}%
  \BibitemOpen
  \bibfield  {author} {\bibinfo {author} {\bibfnamefont {G.~P.}\ \bibnamefont
  {Srivastava}},\ }\href@noop {} {\emph {\bibinfo {title} {The Physics of
  Phonons}}}\ (\bibinfo  {publisher} {Taylor \& Francis Group},\ \bibinfo
  {address} {New York, NY},\ \bibinfo {year} {1990})\BibitemShut {NoStop}%
\bibitem [{\citenamefont {Lindsay}\ \emph
  {et~al.}(2013{\natexlab{b}})\citenamefont {Lindsay}, \citenamefont {Broido},\
  and\ \citenamefont {Reinecke}}]{Lindsay2013PRB}%
  \BibitemOpen
  \bibfield  {author} {\bibinfo {author} {\bibfnamefont {L.}~\bibnamefont
  {Lindsay}}, \bibinfo {author} {\bibfnamefont {D.~A.}\ \bibnamefont {Broido}},
  \ and\ \bibinfo {author} {\bibfnamefont {T.~L.}\ \bibnamefont {Reinecke}},\
  }\href {\doibase 10.1103/PhysRevB.87.165201} {\bibfield  {journal} {\bibinfo
  {journal} {Phys. Rev. B}\ }\textbf {\bibinfo {volume} {87}},\ \bibinfo
  {pages} {165201} (\bibinfo {year} {2013}{\natexlab{b}})}\BibitemShut
  {NoStop}%
\bibitem [{\citenamefont {Mingo}\ \emph {et~al.}(2010)\citenamefont {Mingo},
  \citenamefont {Esfarjani}, \citenamefont {Broido},\ and\ \citenamefont
  {Stewart}}]{Mingo2010}%
  \BibitemOpen
  \bibfield  {author} {\bibinfo {author} {\bibfnamefont {N.}~\bibnamefont
  {Mingo}}, \bibinfo {author} {\bibfnamefont {K.}~\bibnamefont {Esfarjani}},
  \bibinfo {author} {\bibfnamefont {D.~A.}\ \bibnamefont {Broido}}, \ and\
  \bibinfo {author} {\bibfnamefont {D.~A.}\ \bibnamefont {Stewart}},\ }\href
  {\doibase 10.1103/PhysRevB.81.045408} {\bibfield  {journal} {\bibinfo
  {journal} {Phys. Rev. B}\ }\textbf {\bibinfo {volume} {81}},\ \bibinfo
  {pages} {045408} (\bibinfo {year} {2010})}\BibitemShut {NoStop}%
\bibitem [{\citenamefont {Polanco}\ and\ \citenamefont
  {Lindsay}(2018)}]{Polanco2018}%
  \BibitemOpen
  \bibfield  {author} {\bibinfo {author} {\bibfnamefont {C.~A.}\ \bibnamefont
  {Polanco}}\ and\ \bibinfo {author} {\bibfnamefont {L.}~\bibnamefont
  {Lindsay}},\ }\href {\doibase 10.1103/PhysRevB.97.014303} {\bibfield
  {journal} {\bibinfo  {journal} {Phys. Rev. B}\ }\textbf {\bibinfo {volume}
  {97}},\ \bibinfo {pages} {014303} (\bibinfo {year} {2018})}\BibitemShut
  {NoStop}%
\bibitem [{\citenamefont {Katcho}\ \emph {et~al.}(2014)\citenamefont {Katcho},
  \citenamefont {Carrete}, \citenamefont {Li},\ and\ \citenamefont
  {Mingo}}]{Katcho2014}%
  \BibitemOpen
  \bibfield  {author} {\bibinfo {author} {\bibfnamefont {N.~A.}\ \bibnamefont
  {Katcho}}, \bibinfo {author} {\bibfnamefont {J.}~\bibnamefont {Carrete}},
  \bibinfo {author} {\bibfnamefont {W.}~\bibnamefont {Li}}, \ and\ \bibinfo
  {author} {\bibfnamefont {N.}~\bibnamefont {Mingo}},\ }\href {\doibase
  10.1103/PhysRevB.90.094117} {\bibfield  {journal} {\bibinfo  {journal} {Phys.
  Rev. B}\ }\textbf {\bibinfo {volume} {90}},\ \bibinfo {pages} {094117}
  (\bibinfo {year} {2014})}\BibitemShut {NoStop}%
\bibitem [{\citenamefont {Katre}\ \emph {et~al.}(2017)\citenamefont {Katre},
  \citenamefont {Carrete}, \citenamefont {Dongre}, \citenamefont {Madsen},\
  and\ \citenamefont {Mingo}}]{Katre2017}%
  \BibitemOpen
  \bibfield  {author} {\bibinfo {author} {\bibfnamefont {A.}~\bibnamefont
  {Katre}}, \bibinfo {author} {\bibfnamefont {J.}~\bibnamefont {Carrete}},
  \bibinfo {author} {\bibfnamefont {B.}~\bibnamefont {Dongre}}, \bibinfo
  {author} {\bibfnamefont {G.~K.~H.}\ \bibnamefont {Madsen}}, \ and\ \bibinfo
  {author} {\bibfnamefont {N.}~\bibnamefont {Mingo}},\ }\href {\doibase
  10.1103/PhysRevLett.119.075902} {\bibfield  {journal} {\bibinfo  {journal}
  {Phys. Rev. Lett.}\ }\textbf {\bibinfo {volume} {119}},\ \bibinfo {pages}
  {075902} (\bibinfo {year} {2017})}\BibitemShut {NoStop}%
\bibitem [{\citenamefont {Portnoi}\ \emph {et~al.}(1967)\citenamefont
  {Portnoi}, \citenamefont {Romashov}, \citenamefont {Chubarov}, \citenamefont
  {Levinskaya},\ and\ \citenamefont {Salibekov}}]{Portnoi1967}%
  \BibitemOpen
  \bibfield  {author} {\bibinfo {author} {\bibfnamefont {K.~I.}\ \bibnamefont
  {Portnoi}}, \bibinfo {author} {\bibfnamefont {V.~M.}\ \bibnamefont
  {Romashov}}, \bibinfo {author} {\bibfnamefont {V.~M.}\ \bibnamefont
  {Chubarov}}, \bibinfo {author} {\bibfnamefont {M.~K.}\ \bibnamefont
  {Levinskaya}}, \ and\ \bibinfo {author} {\bibfnamefont {S.~E.}\ \bibnamefont
  {Salibekov}},\ }\href {\doibase 10.1007/BF00775639} {\bibfield  {journal}
  {\bibinfo  {journal} {Soviet Powder Metallurgy and Metal Ceramics}\ }\textbf
  {\bibinfo {volume} {6}},\ \bibinfo {pages} {99} (\bibinfo {year}
  {1967})}\BibitemShut {NoStop}%
\bibitem [{\citenamefont {Naslain}\ and\ \citenamefont
  {Kasper}(1970)}]{naslain1970}%
  \BibitemOpen
  \bibfield  {author} {\bibinfo {author} {\bibfnamefont {R.}~\bibnamefont
  {Naslain}}\ and\ \bibinfo {author} {\bibfnamefont {J.~S.}\ \bibnamefont
  {Kasper}},\ }\href@noop {} {\bibfield  {journal} {\bibinfo  {journal}
  {Journal of Solid State Chemistry}\ }\textbf {\bibinfo {volume} {1}},\
  \bibinfo {pages} {150} (\bibinfo {year} {1970})}\BibitemShut {NoStop}%
\bibitem [{\citenamefont {Borgstedt}\ and\ \citenamefont
  {Guminski}(2003)}]{Borgstedt2003}%
  \BibitemOpen
  \bibfield  {author} {\bibinfo {author} {\bibfnamefont {H.~B.}\ \bibnamefont
  {Borgstedt}}\ and\ \bibinfo {author} {\bibfnamefont {C.}~\bibnamefont
  {Guminski}},\ }\href {\doibase 10.1361/105497103772084723} {\bibfield
  {journal} {\bibinfo  {journal} {Journal of Phase Equilibria}\ }\textbf
  {\bibinfo {volume} {24}},\ \bibinfo {pages} {572} (\bibinfo {year}
  {2003})}\BibitemShut {NoStop}%
\bibitem [{\citenamefont {Kang}\ \emph {et~al.}(2017)\citenamefont {Kang},
  \citenamefont {Wu},\ and\ \citenamefont {Hu}}]{Kang2017}%
  \BibitemOpen
  \bibfield  {author} {\bibinfo {author} {\bibfnamefont {J.~S.}\ \bibnamefont
  {Kang}}, \bibinfo {author} {\bibfnamefont {H.}~\bibnamefont {Wu}}, \ and\
  \bibinfo {author} {\bibfnamefont {Y.}~\bibnamefont {Hu}},\ }\href {\doibase
  10.1021/acs.nanolett.7b03437} {\bibfield  {journal} {\bibinfo  {journal}
  {Nano Letters}\ }\textbf {\bibinfo {volume} {17}},\ \bibinfo {pages} {7507}
  (\bibinfo {year} {2017})}\BibitemShut {NoStop}%
\bibitem [{\citenamefont {Whiteley}\ \emph {et~al.}(2011)\citenamefont
  {Whiteley}, \citenamefont {Zhang}, \citenamefont {Gong}, \citenamefont
  {Bakalova}, \citenamefont {Mayo}, \citenamefont {Edgar},\ and\ \citenamefont
  {Kuball}}]{Whiteley2011}%
  \BibitemOpen
  \bibfield  {author} {\bibinfo {author} {\bibfnamefont {C.}~\bibnamefont
  {Whiteley}}, \bibinfo {author} {\bibfnamefont {Y.}~\bibnamefont {Zhang}},
  \bibinfo {author} {\bibfnamefont {Y.}~\bibnamefont {Gong}}, \bibinfo {author}
  {\bibfnamefont {S.}~\bibnamefont {Bakalova}}, \bibinfo {author}
  {\bibfnamefont {A.}~\bibnamefont {Mayo}}, \bibinfo {author} {\bibfnamefont
  {J.}~\bibnamefont {Edgar}}, \ and\ \bibinfo {author} {\bibfnamefont
  {M.}~\bibnamefont {Kuball}},\ }\href {\doibase
  https://doi.org/10.1016/j.jcrysgro.2010.10.057} {\bibfield  {journal}
  {\bibinfo  {journal} {Journal of Crystal Growth}\ }\textbf {\bibinfo {volume}
  {318}},\ \bibinfo {pages} {553} (\bibinfo {year} {2011})}\BibitemShut
  {NoStop}%
\end{thebibliography}

%

\end{document}


\title{Supplementary Material for \\ ``Antisite pairs suppress the thermal conductivity of BAs"}

\author{Qiang Zheng}
\affiliation{Materials Science and Technology Division, Oak Ridge National Laboratory,
Oak Ridge, TN 37831, USA}
\author{Carlos A. Polanco}
\affiliation{Materials Science and Technology Division, Oak Ridge National Laboratory,
Oak Ridge, TN 37831, USA}
\author{Mao-Hua Du}
\affiliation{Materials Science and Technology Division, Oak Ridge National Laboratory,
Oak Ridge, TN 37831, USA}
\author{Lucas R. Lindsay}
\affiliation{Materials Science and Technology Division, Oak Ridge National Laboratory,
Oak Ridge, TN 37831, USA}
\author{Miaofang Chi}
\affiliation{Center for Nanophase Materials Sciences, Oak Ridge National Laboratory,
Oak Ridge, TN 37831, USA}
\author{Jiaqiang Yan}
\affiliation{Materials Science and Technology Division, Oak Ridge National Laboratory,
Oak Ridge, TN 37831, USA}
\affiliation{Department of Materials Science and Engineering, University of Tennessee, Knoxville, TN 37996, USA}
\author{Brian C. Sales}
\affiliation{Materials Science and Technology Division, Oak Ridge National Laboratory,
Oak Ridge, TN 37831, USA}

\date{\today}

\maketitle

\makeatletter
\renewcommand{\thefigure}{S\@arabic\c@figure}
\newcommand{\tm}{\textit{T\hspace{-0.25ex}M}}
\makeatother

\section{SEM \MakeLowercase{and} EDX \MakeLowercase{for} \MakeLowercase{a} BA\MakeLowercase{s} \MakeLowercase{single crystal}}

\begin{figure}[b]
\includegraphics[width=0.32\textwidth]{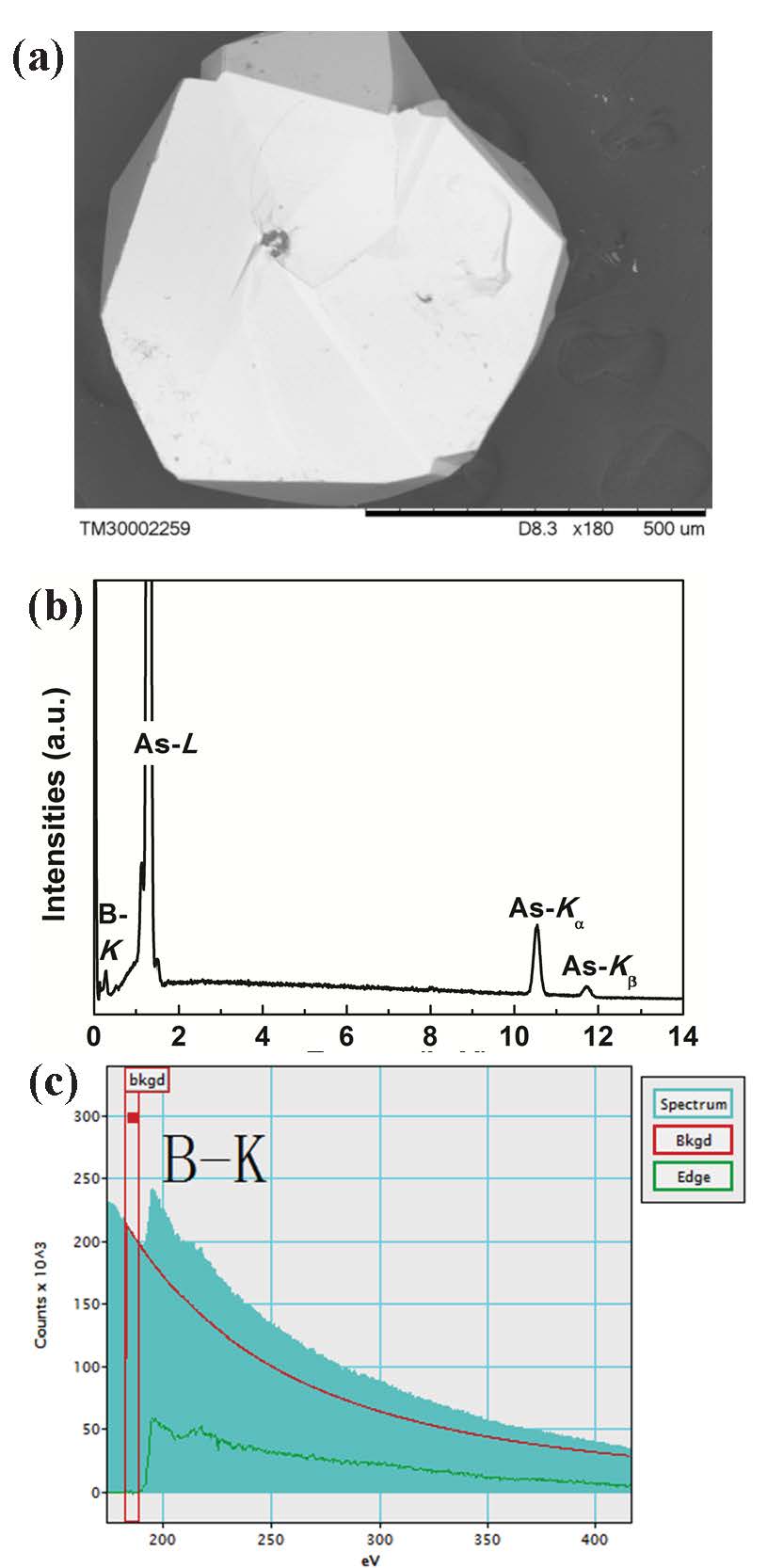}
\caption{(Color online) (a) A scanning electron microscopy (SEM) image for a BAs single crystal. (b) An energy-dispersive X-ray spectrum (EDX) for this BAs single crystal,
indicating besides As and B, no other element was detected. (c) An EEL spectrum for a region with
As$_\mathrm{B}$-B$_\mathrm{As}$ pairs. We also collected spectra covering the energy-loss ranges of I, O, Si and Al, showing no sign of these elements.}
\label{SEM_BAs}
\end{figure}

A scanning electron microscopy (SEM) image for a BAs single crystal is displayed in Fig.\ \ref{SEM_BAs}(a),
and its corresponding energy-dispersive X-ray spectrum (EDX) is given in Fig.\ \ref{SEM_BAs}(b).
No foreign atoms are observed in EDX.

\section{HAADF \MakeLowercase{imaging simulation}}

HAADF images along [001] for the perfect structure and structures with As$_\mathrm{B}$, B$_\mathrm{As}$ and $V_\mathrm{As}$ were simulated within the QSTEM program package \cite{koch2008qstem}. Since electron probe intensity oscillates when electrons enter into a crystal matrix, probe channelling, $i.e.$ the tendency of electron probe to remain in an atomic column, would be controlled by different configurations of atoms in each column \cite{esser2016,Hwang2016,Haruta2009}. Therefore, during the image simulation, this probe channeling due to different configuration of defect atoms in each column was taken into account.

\begin{figure}
\includegraphics[width=0.4\textwidth]{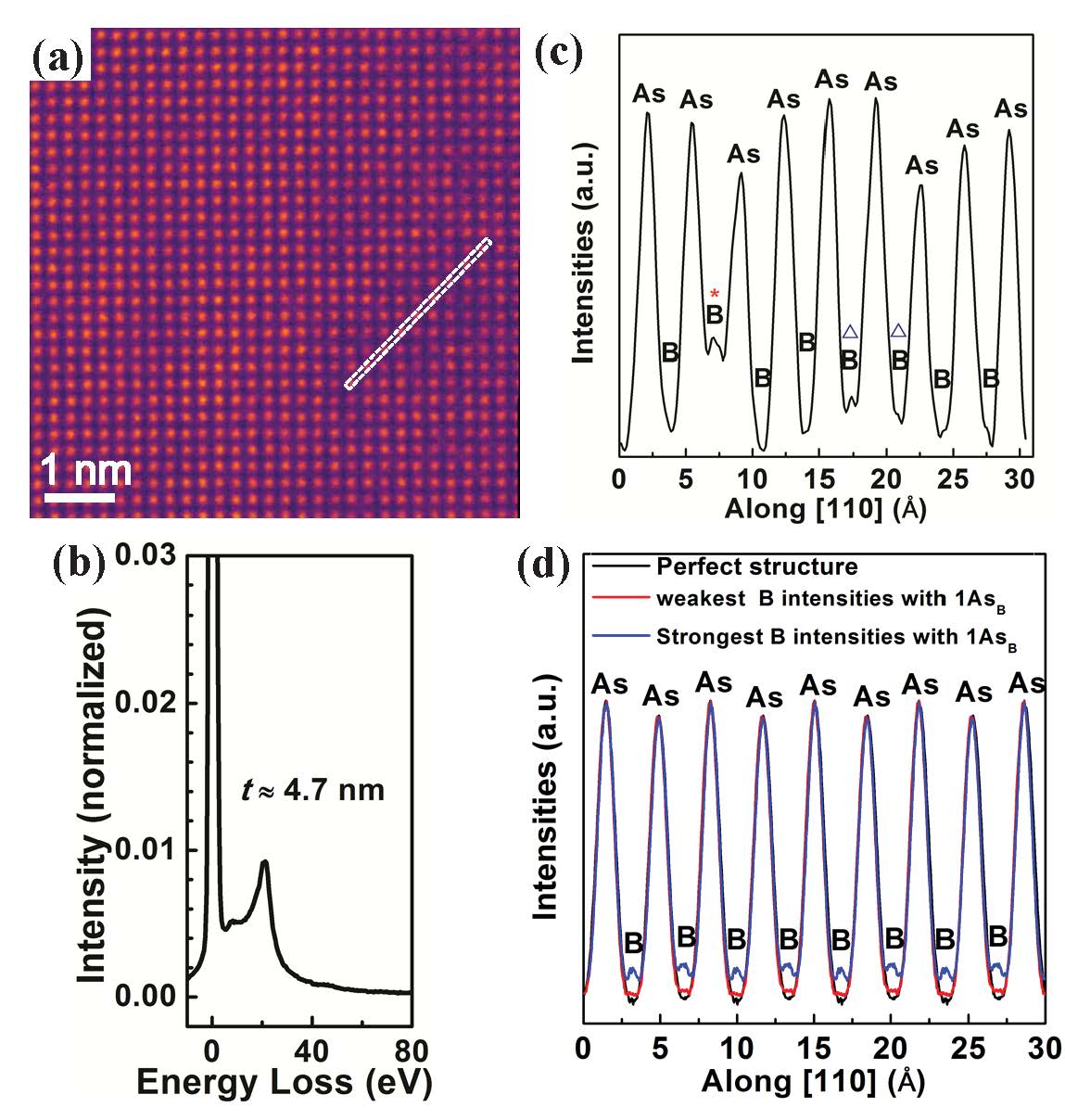}
\caption{(Color online) (a) A HAADF image along [001] for a region with thickness of 4.7 \,nm ($\sim$10 unit cells). The thickness for this region was calculated from its EEL spectrum in (b).
The intensity profile for rectangle region in (a) is displayed in (c), confirming the existence As$_\mathrm{B}$ antisite
defects.
The B columns marked by blue triangles in (c) most likely contain 1As$_\mathrm{B}$ in each of them, while the one marked by red asterisk indicates more than one As$_\mathrm{B}$ in it.
(d) HAADF images along [001] for the perfect BAs structure with thickness of 10$a$
and the structures with 1As$_\mathrm{B}$ defect in each B column are simulated.
The probe channeling was taken into account, ${i.e.}$ 10 possibilities of 1As$_\mathrm{B}$ in each B column were simulated.
The comparison among simulated intensity profiles for the columns along [110] in the perfect structure and the structures with 1As$_\mathrm{B}$ is shown.
Note that only maximum and minimum simulated intensities for B columns with 1As$_\mathrm{B}$ are displayed.}
\label{STEM_10a_simulated}
\end{figure}

For the region with thickness of  10$a$,  only one As$_\mathrm{B}$ antisite in a B column was simulated.
Since probe channeling could result in obvious intensity variation between different configurations for a given composition \cite{esser2016},
all 10 possibilities of 1As$_\mathrm{B}$ in a B column were simulated. Fig.\ \ref{STEM_10a_simulated}(d) demonstrates the comparison among simulated intensity profiles for atomic columns along [110]
in the perfect structure and the structures with 1As$_\mathrm{B}$ in each B column.
Only maximum and minimum simulated intensities of B columns are shown. The intensities of most visible B columns in Fig.\ \ref{STEM_10a_simulated}(a) are between the maximum and minimum, suggesting each of them only contains 1As$_\mathrm{B}$. However, as reflected by the B column marked by red asterisk in Fig.\ \ref{STEM_10a_simulated}(c),
some visible B columns show much stronger intensities than the maximum simulated intensity for a B column with 1As$_\mathrm{B}$,
indicating each of them contains more than one As$_\mathrm{B}$. This is reasonable, since at a thicker region,
the probability of a B column involving more than one As$_\mathrm{B}$ is higher.

\begin{figure} [!t]
\includegraphics[width=0.38\textwidth]{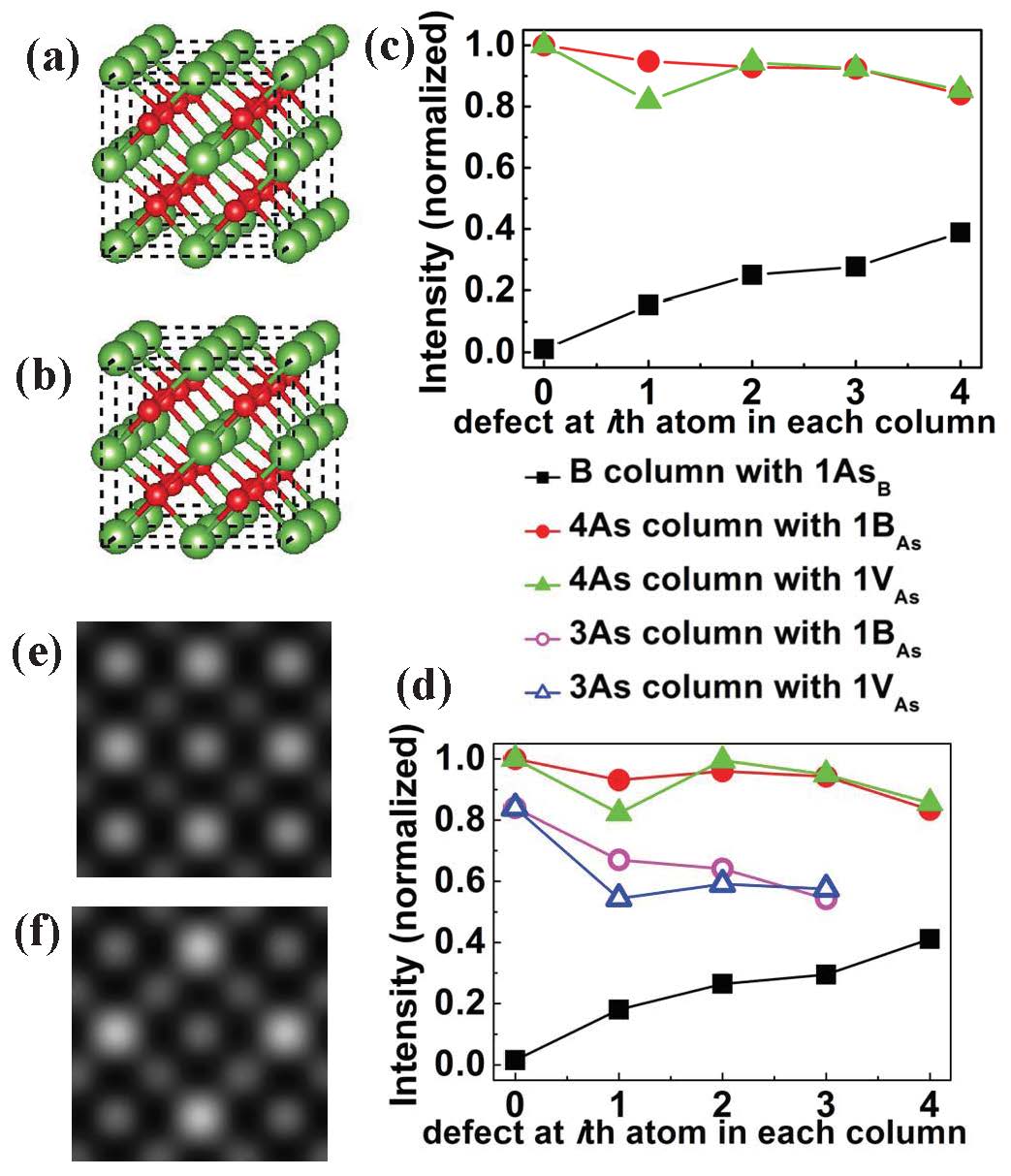}
\caption{(Color online)
Structure model with thickness of (a) 3.75$a$ and (b) 3.5$a$ was used for the HAADF image simulations along [001], respectively.
Except each of the As columns at (0,0,$z$) and ($\frac{1}{2}$,$\frac{1}{2}$,$z$) in the latter one contains 3 atoms, all other columns
in the two models involve 4 atoms for each. The intensity variations for an As column with 1B$_\mathrm{As}$ or 1V$_\mathrm{As}$ defect
and a B column with 1As$_\mathrm{B}$ defect in the 3.75$a$ and 3.5$a$ models are demonstrated in (c) and (d), respectively.
All intensities were normalized with respect to the intensity of the perfect As column involving 4As. The probe channeling was taken into account,
as revealed by $i$th atom in beam incident direction substituted by one defect. $i$ = 0 indicates the intensities in the perfect structures.
The simulated HAADF images for the 3.75$a$ and 3.5$a$ model with one antisite pair, which result in strongest B intensities and weakest As intensities
due to probe channeling, are displayed in (e) and (f), respectively.}
\label{STEM_3.75a_simulated}
\end{figure}

HAADF image simulations were also performed to better visualize the intensity variations caused by possible
As$_\mathrm{B}$ in B columns and B$_\mathrm{As}$ or V$_\mathrm{As}$ in As columns for the region with thickness of 3.6$a$.
Since no model could be built for the thickness of 3.6$a$, we then built two models with thicknesses most close to 3.6$a$: one is 3.75$a$ and the other is 3.5$a$, as shown in Fig.\ \ref{STEM_3.75a_simulated}(a) and (b), respectively. Except each of the As
columns at (0,0,$z$) or ($\frac{1}{2}$,$\frac{1}{2}$,$z$) in the latter one contains 3 atoms, each of other columns in the two models involves 4 atoms.

Fig.\ \ref{STEM_3.75a_simulated}(c) displays the intensity variations of a B column with 1As$_\mathrm{B}$ and an As column with 1B$_\mathrm{As}$ or
1V$_\mathrm{As}$ for the 3.75$a$ model. All intensities were normalized with respect to the intensity of an As column involving 4As.
The intensity of a B column with 1As$_\mathrm{B}$ varies in a wide range from 0.18 to 0.4 due to the different configurations of
1As$_\mathrm{B}$ in it. 1B$_\mathrm{As}$ or 1V$_\mathrm{As}$ in an As column causes nearly the same intensity weakening for this column in the range of 0.81--0.94.

Intensity variations for the 3.5$a$ model
with different types of defects are shown in Fig.\ \ref{STEM_3.75a_simulated}(d). The intensity variations of a B column with 1As$_\mathrm{B}$ and
a 4As column with 1B$_\mathrm{As}$ or 1V$_\mathrm{As}$ are similar to those in the 3.75$a$ model. Interestingly,
the intensity of a 3As column  could drop from 0.87 to 0.55 due to 1B$_\mathrm{As}$ or 1V$_\mathrm{As}$
defect in it, of which the minimum is only a little bit stronger than the maximum intensity of a B column with 1As$_\mathrm{B}$.

The corresponding simulated images with
strongest B intensities and weakest As intensities due to probe channeling for the two models with one antisite pair are given in Fig.\ \ref{STEM_3.75a_simulated}(e)
for the 3.75$a$ model and (f) for the 3.5$a$ model, consistent with the HAADF image in Fig.\ 1(b).

\begin{figure} []
\includegraphics[width=0.32\textwidth]{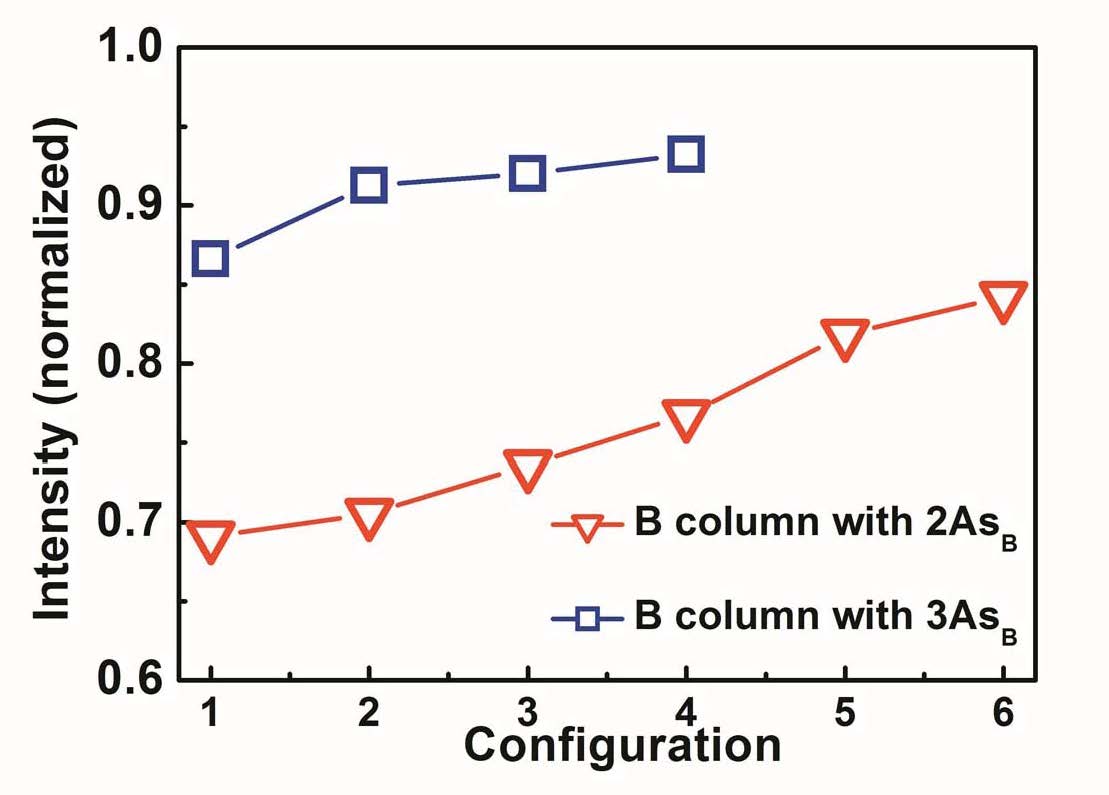}
\caption{(Color online)
The intensity variations for a B column with 2As$_\mathrm{B}$ or 3As$_\mathrm{B}$ in the 3.75$a$ model taking into account the possible As$_\mathrm{B}$ configurations in each case.
All intensities were normalized with respect to the intensity of the perfect As column involving 4As.
Results have been ordered from the smallest to the largest intensities. The intensity variations of a B column with 2As$_\mathrm{B}$ or 3As$_\mathrm{B}$ in the 3.5$a$ model are almost same with those in the 3.75$a$ model.}
\label{STEM_2and3As_B_simulated}
\end{figure}

Finally, to assess if each B column with visible intensity in Fig.\ 1(b) could involve more As$_\mathrm{B}$ defects,
intensity variations for a B column with 2As$_\mathrm{B}$ or 3As$_\mathrm{B}$ were simulated for both models,
as those for the 3.75$a$ model displayed in Fig.\ \ref{STEM_2and3As_B_simulated}.
The intensities of a B column with 2As$_\mathrm{B}$ and 3As$_\mathrm{B}$ vary in the range of 0.69-- 0.84 and 0.87--0.93, respectively.
2As$_\mathrm{B}$ or 3As$_\mathrm{B}$ in a B column in the 3.5$a$ model reveal almost same intensity variations.
Such strengthened intensities for a B column with 2As$_\mathrm{B}$ or 3As$_\mathrm{B}$ are much stronger than those for a B column with 1As$_\mathrm{B}$ (0.18--0.4),
suggesting almost all visible B columns in Fig.\ 1(b) only contain 1As$_\mathrm{B}$ for each.

\section{C\MakeLowercase{omputational Methods for formation energies of defects}}

Our calculations are based on density functional theory (DFT) implemented in the plane-wave basis VASP code \cite{KRESSE1996A}.
The projector augmented wave method was used to describe the interaction between ions and electrons \cite{KRESSE1999}. The kinetic energy cutoff is 319\,eV. A 216-atom supercell was used for defect calculations.
The reciprocal-space integrations were performed on a 2$\times$2$\times$2 $k$-point mesh.
Structures and total energies of defects were obtained by DFT calculations using Perdew, Burke, and Ernzerhof (PBE) \cite{Perdew1996} functionals
while the band gap was corrected using Heyd-Scuseria-Ernzerhof (HSE) hybrid functional \cite{Heyd2003,Krukau2006},
which includes 25\,\% Hartree-Fock exchange \cite{Biswas2012}.
The optimized lattice constant of BAs (4.818\,\AA) is in good agreement with the experimental value of 4.7776\,\AA \cite{Perri1958}.
The atomic positions were relaxed until the residual forces were less than 0.02\,eV/\AA.
The calculated indirect band gap of BAs is 1.20\,eV at the PBE level and 1.62\,eV at the HSE level.

The defect formation energy $\Delta$$H$ was calculated according to
\begin{equation}
{\Delta}H = (E_D-E_0)-\sum_{i}n_i(\mu_i+\mu_i^{bulk})+q(\varepsilon_{VBM}+\varepsilon_f)+{\Delta}E_{corr}
\label{energy}
\end{equation}
In the first term $E_D$ and $E_0$ are the total energies of the defect-containing and the defect-free supercells, respectively. The second term in Eq.\ \ref{energy} is the change in energy due to the exchange of atoms with their reservoirs,
where $n_i$ is the difference in the number of atoms for the $i$th atomic species between the defect-containing and defect-free supercells.
$\mu_i$ is the chemical potential of the $i$th atomic species relative to its bulk chemical potential $\mu_i^{bulk}$. The third term is the change in energy due to the exchange of electrons with their reservoir.
$q$ is the defect charge state. $\varepsilon_{VBM}$ is the energy of the valence band maximum (VBM) of BAs.
$\varepsilon_f$ is the Fermi energy relative to the VBM and can be varied between the VBM and the conduction band minimum (CBM).
The fourth term includes potential alignment and image charge corrections \cite{Lany2008}.

The chemical potentials in Eq.\ \ref{energy} are subject to constraints under thermal equilibrium. To maintain the stability of BAs without bulk B and As precipitation during synthesis, the chemical potentials of B and As should satisfy the following:
\begin{equation}
\mu_\mathrm{B}+\mu_\mathrm{As}={\Delta}H(\mathrm{BAs})
\label{eq2}
\end{equation}
\begin{equation}
\mu_\mathrm{B}\leq0, \mu_\mathrm{As}\leq0
\label{eq3}
\end{equation}
${\Delta}H$(BAs) is the enthalpy of formation for BAs, which is calculated to be -0.07\,eV. Since BAs was synthesized under the As rich condition,
$\mu_\mathrm{As}$ = 0 was chosen for the defect calculations. Note that the small ${\Delta}H$($\mathrm{BAs}$) leads to a small range of $\mu_\mathrm{As}$ (-0.07\,eV$\leq$$\mu_\mathrm{As}$$\leq$0 ).
Varying $\mu_\mathrm{As}$ within its allowed range under thermal equilibrium does not change the defect formation energies significantly.
To prevent the formation of B$_6$As phase, the following condition should also be satisfied:
\begin{equation}
\mu_\mathrm{B}+\mu_\mathrm{As}\leq{\Delta}H(\mathrm{B_6As})
\label{eq4}
\end{equation}
Here, ${\Delta}H$(B$_6$As) is the enthalpy of formation for B$_6$As, which is calculated to be -1.35\,eV.
Eq.\ \ref{eq4} leads to $\mu_\mathrm{B}$+$\mu_\mathrm{As}\geq$0.19\,eV.
Thus, Eq.\ \ref{eq3}  and Eq.\ \ref{eq4}  cannot be simultaneously satisfied. These calculations suggest that BAs is unstable against segregation into the boron-rich phase of B$_6$As and bulk As,
in agreement with a previous DFT study \cite{Ektarawong2017}.
However, the B$_6$As phase was not observed in BAs crystals synthesized in this work. This could be due to the effect of entropy.
It is also likely that the thermal equilibrium is not reached due to the insufficient atomic diffusion at the growth temperature,
which prevents the formation of B$_6$As that has a very different crystal structure from BAs.

The defect concentration ($N$) at thermal equilibrium can be calculated by
\begin{equation}
N = N_\mathrm{0}\mathrm{exp}(-{\Delta}H/k_BT)
\label{eq5}
\end{equation}
where $N_\mathrm{0}$ is the number of the available sites for defect formation, $k_B$ is Boltzmann constant, and $T$ is temperature and $\Delta$$H$ is the defect formation energy calculated by Eq.\ \ref{energy}.

\section{T\MakeLowercase{hermal conductivity calculations}}

The thermal conductivity of BAs was calculated solving the Peierls-Boltzmann transport equation self-consistently \cite{Ziman2001,Srivastava1990,Lindsay2013PRB},
including thermal resistance from three-phonon \cite{Ziman2001,Srivastava1990} and phonon-defect scattering.
Specifically, phonon-defect scattering is computed using a parameter-free $ab$ initio Green’s function methodology \cite{Mingo2010,Protik2016PRB,Polanco2018},
which includes the interatomic force constant (IFC) variance locally near defects. The IFCs required for our framework are calculated using finite differences of energies extracted from density functional theory using Quantum Espresso \cite{Giannozzi2009}.
For each energy calculation we use norm conserving pseudopotentials within the local density approximation and Perdew-Zunger parametrization \cite{Perdew1981}.
The IFCs locally near defects were calculated using a similar methodology after a 432-atom supercell containing the defect was relaxed until the interatomic forces are less than 10$^{-5}$\,Ry/bohr. We include IFCs up to the 10$^\mathrm{th}$ atomic shell and enforce point group symmetry as well as translational symmetry using the procedure outlined by Esfarjani $et$ $al.$ \cite{Esfarjani2008}.


%